\documentclass[twocolumn,aps,showpacs,nofootinbib,epsfig]{revtex4}
\usepackage{axodraw}
\usepackage{epsfig}
\usepackage{amsmath}
\usepackage{graphicx}
\usepackage{amssymb}
\newcommand{\be}{\begin{equation}}
\newcommand{\ee}{\end{equation}}
\newcommand{\bea}{\begin{eqnarray}}
\newcommand{\eea}{\end{eqnarray}}
\newcommand{\beas}{\begin{eqnarray*}}
\newcommand{\eeas}{\end{eqnarray*}}

\newcommand{\M}{\mathcal{M}}

\newcommand{\qv}{ \vec{Q}_{\parallel}}

\begin{document}
\title{Radiative and non-perturbative  corrections to  the electron mass  and the anomalous magnetic moment in the presence of an external magnetic field of arbitrary strength}
\author{Eduardo Rojas Pe\~na}    
\affiliation{Instituto de Ciencias Nucleares, Universidad
Nacional Aut\'onoma de M\'exico, Apartado Postal 70-543, M\'exico
Distrito Federal 04510, M\'exico}
\begin{abstract}
Using  the Ritus eingenfunction method we   compute  corrections to the electron mass $m_{0}$  in the presence of a moderate magnetic field   $eB\sim m_0^2$. From this we obtain an expression for the anomalous magnetic moment near  the critic field. For this we solved numerically  the Schwinger-Dyson equations in the rainbow approximation  including all Landau levels  without   make any assumption respect to the field strength.
\end{abstract}

\maketitle

\section{Introduction}\label{sec1}
Radiative corrections to the electron self-energy  due  to its interaction with an external magnetic field have been widely considered in the literature~\cite{demeur,newton,jancovici,yu,gepraegs,gusyninreal,kuznetsov,wang}.
The exact form of how the ground state energy of an electron may be shifted from $E_0$ in the presence of an magnetic field was a source of  misleading conclusions~\cite{jancovici,connell,duncan2}. Assuming that this effects can be calculated by adding to the Dirac Hamiltonian a term $\Delta\mu {\bf \sigma\cdot B}$, where $\Delta\mu$ is the Schwinger anomalous magnetic moment, it was founded  that the ground-state energy of an electron is~\cite{jancovici,johnson,ternov} 
\begin{align}\label{debil}
E_{0}=m_{0}c^2\lvert 1-\frac{\alpha}{4\pi}\frac{e\hbar B}{m^2_{0}c^3} \rvert,
\end{align}
where $m_0$ is the electron mass without magnetic fields. For $B\sim 10^{16}\text{Gauss}$ this formula implies that the ground-state energy of an electron is close to zero,  a fact that could have dramatical astrophysical and cosmological consequences~\cite{connell}. However, it was soon  realized that this result can only be true for  magnetic field intensities that  are  weak compared  with $B_c = m_0^2/e\sim 10^{13}\text{Gauss}$, because Eq~(\ref{debil}) only represents the linear term in the power expansion in   $eB$~\cite{jancovici}, which is not the dominant one for  strong magnetic fields.  In  recent publications~\cite{wang,klimenko} dynamical  effects of a  strong magnetic field $B>> m_{0}^2/e$  were calculated. These approaches, show an enhancement of the dynamical contribution to the dynamical mass. They  claim that this is a new effect that could have important  consequences. We  confirmed this result with  an independent calculation and  show  that this effect is well described by previous calculations  which are based on the  constant mass approximation~\cite{gusyninreal}. Furthermore, we carry out an analogous calculation for a weak  magnetic  field  $ eB\lesssim m_{0}^2$, from this we  obtain an expression for the anomalous magnetic moment near the critic field $B_c$.
Recently this issue  has been  revived  from a non-perturbative point of view  for strong fields~\cite{Ferrer4}.~
In  our approach we have included higher Landau levels to allow magnetic field intensities of the same order or less  than the electron mass using similar techniques to those used in  the context of the magnetic catalysis~\cite{Gusynin,Leung,Leung3,Ferrer,fraga,ayala,rojas}, particularly we follow the last reference.
\section{The mass function}\label{sec2}
It has been shown~\cite{Ritus} that the
mass operator in the  presence of an electromagnetic field can be written as a
combination of the structures
\be
   \gamma^\mu\Pi_\mu\,,\, \sigma^{\mu\nu}F_{\mu\nu}\,,\,
   (F_{\mu\nu}\Pi^\nu)^2\,,\,\gamma_5F_{\mu\nu}\tilde{F}^{\mu\nu}
   \label{structures}
\ee
which commute with the operator $(\gamma\cdot\Pi)^2$, where 
$
   \Pi_\mu = i\partial_\mu - eA^{\mbox{\tiny{ext}}}_\mu,\
   F_{\mu\nu}=\partial_\mu A^{\mbox{\tiny{ext}}}_\nu -
              \partial_\nu A^{\mbox{\tiny{ext}}}_\mu,\
   \tilde{F}^{\mu\nu}=\frac{1}{2}\epsilon^{\mu\nu\lambda\tau}
   F_{\lambda\tau},\
   \sigma_{\mu\nu}=\frac{i}{2}[\gamma_\mu , \gamma_\nu]
$
and $A^{\mbox{\tiny{ext}}}$ is the external vector
potential. We take 
$A^{\mbox{\tiny{ext}}}_\mu=B(0,-y/2,x/2,0)$, such that the magnetic field is
${\mathbf{B}}=B{\mathbf{\hat{z}}}$.
The equation relating the two-point fermion Green's function $G(x,y)$  and the
mass operator $M(x,y)$ in coordinate space reads
\bea
   \gamma\cdot\Pi(x)G(x,y)\!-\!\!\int \!\!d^4x'M(x,x')G(x',y)
   =\delta^4(x-y),
   \label{Greeneq}
\eea

\noindent where the mass function in the rainbow approximation in coordinate
space is~\cite{Schwinger} 
\begin{align}
   M(x,x')=&m_{0}\delta^{4}(x-x^{\prime})1_{1\times 1}\notag\\
&-ie^2\gamma^\mu G(x,x')\gamma^\nu D^{(0)}_{\mu\nu}(x-x')\, ,
 \label{SDcoord}
\end{align}
where $1_{4\times 4}$ is the $4\times 4$ spinorial identity matrix.
In this work, we are interested in exploring the
behavior of the mass function for arbitrary values of the magnetic field
strength, thus we  work near the Landau
gauge~\cite{miransky} ($\xi\sim 0$) to compare with the well known results  in the  weak and strong magnetic field limit. 

\begin{figure}[htb]
\includegraphics[bb=0  0  15cm 6cm]{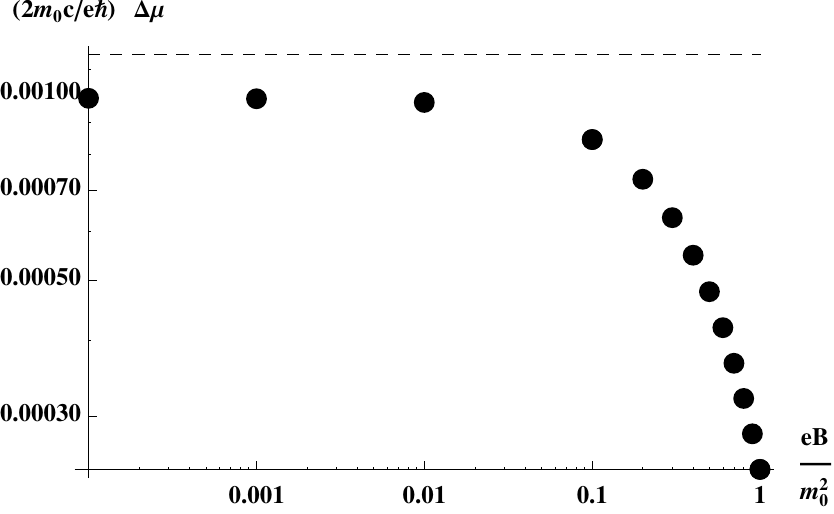}
\caption{Anomalous magnetic moment $\Delta\mu$ as a function of the magnetic field in units of $m_0^2$. Dots   
correspond to the anomalous magnetic moment $\Delta\mu$ taking into account all Landau levels. Horizontal dashed line  corresponds to Schwinger classical  result $\frac{2m_0c\Delta\mu}{e_0\hbar}=\frac{\alpha }{2\pi }$ }
\label{moment}
\end{figure}

Schwinger~\cite{Schwinger} was  the first  to obtain an exact analytical
expression  for  the fermion Green's function in the  presence  of  a constant
electromagnetic field  of  arbitrary  strength. However, in the presence  of  a constant
external field,  the fermion  asymptotic  states  are no   longer  free
particle  states, but  instead are described by eigenfunctions of the operator
$(\gamma^{\mu}\Pi_{\mu})^{2}$. Thus, the alternative representation of $G(x,y)$  proposed  by Ritus~\cite{Ritus} is more
convenient  for our  purposes, since there   the  mass operator  is diagonal. A matrix constructed out of these eigenfunctions is
used to  {\it rotate} the SDE (\ref{SDcoord})  to momentum  space, yielding 
\cite{ayala} 
\begin{widetext}
\begin{align}\label{fulleq}
   \text{tr}[\Pi(n_p)]\left(\mathcal{M}[p_\parallel,n_p]-m_{0}\right)=&-2ie^2\sum_{\sigma_k,\sigma_p=\pm
   1} 
   \sum_{n_k,s_k=0}^{\infty}\frac{s_k! (n_p-\frac{(\sigma_p+1)}{2})!}{s_p!
   (n_k-\frac{(\sigma_k+1)}{2})!} \nonumber\\
   \times &
   \int \frac{d^4q}{(2\pi)^4}\frac{e^{-q_\perp^2/2\gamma}}{q^2+i\epsilon}
   \frac{{\mathcal{M}}[(p-q)_\parallel,n_k]}{(p-q)_\parallel^2-
   2eBn_k-{\mathcal{M}}^2[(p-q)_\parallel,n_k]}
   \nonumber\\
      \times\left(\frac{q_\perp^2}{4\gamma}\right)^
   {l_k-l_p-\frac{(\sigma_k-\sigma_p)}{2}}\!\!
   &\left[ 2  + \frac{(1-\xi)}{q^2}\left(q_\perp^2(1-\delta_{\sigma_p\sigma_k})
   -q_\parallel^2\delta_{\sigma_p\sigma_k}\right)\right]\!\! 
   \left[L_{n_p-\frac{(\sigma_p+1)}{2}}^{n_k-n_p-
   \frac{(\sigma_k-\sigma_p)}{2}} 
   (q_\perp^2/4\gamma)\right]^2\!\!\!\!
   \left[L_{s_k}^{s_p-s_k}(q_\perp^2/4\gamma)\right]^2
\end{align}
\end{widetext}
where $q_\perp = (0,q_1,q_2,0)$,  $q_\parallel=(q_0,0,0,q_3)$,  $\gamma=eB/2$,
$p^2=E_p^2-p_z^2-2eBn$ and $L_{n}^{m}$ are  Laguerre  functions and
\begin{align}
\Pi(n_p)=
\begin{cases}
1,  & \text{if $n_p\ne 0$}\notag\\
\Delta(\sigma=1), & \text{if $n_p = 0$},
\end{cases},
\end{align}
with $\Delta(\sigma)=\frac{1}{2}(1+i\sigma\gamma^1\gamma^2)$. This factor comes from the normalization~\cite{Leung3}
\begin{align}
\int d^{4}x\bar{\Psi}_p(x)\Psi_{p^{\prime}}(x)=\delta_{n_p n_{p^{\prime}}}\delta_{s_p s_{p^{\prime}}}\delta^2(p_{\parallel}-p^{\prime}_{\parallel})\Pi(n_p),
\end{align}
and bring out a factor of two in the left-hand side of Eq.~(\ref{fulleq}) for $n_p=0$.
In vacuum, working near the Landau gauge, we know that the wave function
renormalization equals one. We  assume that this is the case
also in the presence of a magnetic field of arbitrary strength.
Accordingly, to get  Eq.~(\ref{fulleq}) we have worked with the ansatz $\Sigma(\bar{p})\sim {\mathcal{M}}(\bar{p})1_{4\times 4}$, 
where $\Sigma(\bar{p})$ is defined by
\begin{align}\label{apend42}
\int d^4x & d^4x^{\prime}\bar{\Psi}_p(x)M(x,x^{\prime})\Psi_{p^{\prime}}(x^{\prime})\notag\\
=&\delta_{n_p n_{p^{\prime}}}\delta_{s_{p}s_{p^{\prime}}}\delta^{2}(p_{\parallel}-p_{\parallel}^{\prime})\Pi(n_{p})\Sigma(\bar{p}).
\end{align}
This assumption is good when considering   a small external momentum and  $n_p=0$~\cite{Leung,ayala,rojas,Ferrer4}.
Since the energy only depends on the
principal quantum number $n_k$, we expect that ${\mathcal{M}}((p-q)_\parallel,n_k)$ should be 
independent of $s_k$. We also assume that
${\mathcal{M}}((p-q)_\parallel,n_k)$ is a slowly varying function of
$n_k$ and thus make the approximation
\begin{equation}\label{aproximacion}
{\mathcal{M}}((p-q)_\parallel,n_k)\sim{\mathcal{M}}((p-q)_\parallel,n_k=0).
\end{equation}
 Hereafter, we employ the
more convenient notation
${\mathcal{M}}(k_\parallel,n_k=0)\equiv{\mathcal{M}}(k_\parallel)$ 
for generic arguments of the mass function. With these
considerations, the sum over $s_k$ can be computed by using the identity
\begin{equation}
   L_{s}^{l}=(-1)^{l}x^{-l}\frac{(s+l)!}{s!}L_{s+l}^{-l},
   \label{ident1}
\end{equation}
and  the Eq~5.11.5.1 in Ref.~\cite{prudnikov}, namely
\begin{align}
   \sum_{k=0}^{\infty}A^{k}L_{k}^{m-k}(x)L_{n}^{k-n}(y)=&A^{m}
   \left(\frac{1+A}{A}\right)^{m-n}e^{-xA}\nonumber\\
   \times&L_{m-n}^{0}(x+y+xA+\frac{y}{A}),\nonumber\\
   \label{gaut}
\end{align}
yielding
\begin{align}
\text{tr}&[\Pi(n_p)]\left(\M(p)-m_{0}\right)=-2ie^2\sum_{\sigma_{p},\sigma_{k}=\pm 1}\sum_{k=0}^{\infty}\int
\frac{d^4q}{(2\pi)^4}\nonumber\\
\times &\frac{e^{-\frac{q_{\perp}^2}{4\gamma}}}{q^2+i\varepsilon}
\frac{\M((p-q)_\parallel)}{\left\{(q-p)_\parallel^2-
    2eB(k+(\sigma_{k}+1)/2)\right\}-\M^2}\nonumber\\  
\times&\left\{2+(1-\xi)(1-\delta_{\sigma_{p}\sigma_{k}})\frac{q_{\perp}^2}{q^2}
-(1-\xi)\delta_{\sigma_{p}\sigma_{k}}\frac{q_{\parallel}^2}{q^2}\right\}
\nonumber\\    
\times&(-1)^{-m}(-1)^kL_{m}^{k-m}(\frac{q_{\perp}^2}{4\gamma})
L_{k}^{-(k-m)}(\frac{q_{\perp}^2}{4\gamma}), 
\end{align}
where $k=n_{k}-\frac{\sigma_k+1}{2}$ and $m=n_{p}-\frac{\sigma_p+1}{2}$. It is worth mentioning that after
summing over $s_k$, the resulting equation is the same as Eq.~(50) in
Ref.~\cite{Leung} when considering the case $n_k=0$, which corresponds to the
strong field limit. Under a Wick rotation we have
\begin{align}
   &\frac{1}{(p-q)_\parallel^2-2eB(k+\frac{\sigma_k+1}{2})-
   {\mathcal{M}}^2}\longrightarrow\nonumber\\
   &\frac{-1}{(p-q)_\parallel^2+2eB(k+\frac{\sigma_k+1}{2})+
   {\mathcal{M}}^2}\nonumber\\
=&\frac{-1}{2eB}\int_{0}^{1}dx x^{\left((p-q)_\parallel^2+2eB(k+\frac{\sigma_k+1}{2})+
   {\mathcal{M}}^2-2eB\right)/2eB}.
   \label{denomitor}
\end{align}
With this result the sum over $k$ can be performed also by resorting to
Eq.~(\ref{gaut}) yielding, after carrying out the sums over $\sigma_k$ and
$\sigma_p$,  
%
\begin{align}
   \text{tr}&[\Pi(n_p)]\left(\M(p)-m_{0}\right)=+2e^2\int\frac{d^4Q}{(2\pi)^4}
   \frac{\M (\frac{p}{4\gamma}-Q)_\parallel}{Q^2}\nonumber\\
   &\int_{0}^{1}dx
   e^{-Q_{\perp}^2\left[1-x\right]} 
   x^{\left[(2\sqrt{\gamma}Q-p)_\parallel^2 
   +\M^2\right]\frac{1}{4\gamma}}\nonumber\\ 
   \times&\Biggl[\left\{2-(1-\xi)\frac{Q_{\parallel}^2}{Q^2}\right\}
 \left(x^{n_p}L_{n_p}^{0}+x^{n_p}L_{n_p-1}^{0}\right)\notag\\
   +&\left\{2-(1-\xi)\frac{Q_{\perp}^2}{Q^2}\right\}
\left(x^{(n_p+1)}L_{n_p}^{0}+x^{(n_p-1)}
   L_{n_p-1}^{0}\right)\Biggl].\notag\\
   \label{minkowskytoeucledian}
\end{align}
where $Q=\frac{q}{2\sqrt{\gamma}}$ and 
the argument of
the Laguerre functions is $4Q^2_{\perp}\sin \text{Ln}(\frac{1}{x})$.
%
 For consistency with the assumption that the mass
function is a slowly varying function of the principal quantum number Eq.~(\ref{aproximacion}), we take
$n_p=0$. In this case, Eq.~(\ref{minkowskytoeucledian}) gets simplified since
$L^{0}_{0}=1$ and $L^{0}_{-1}=0$, so we get
\begin{widetext}
\begin{align}
   \M (p)=&m_{0}+e^2\int\frac{d^4Q}{(2\pi)^4}
   \frac{\M (\frac{p}{4\gamma}-Q)_\parallel} {Q^2}\int_{0}^{1}
   dxe^{-Q_{\perp}^2\left(1-x\right)}
    \ x^{\left[(Q-p)_\parallel^2+\frac{\M^2}{2eB}-1\right]}
   \Biggl[\left\{2-(1-\xi)\frac{Q_{\parallel}^2}{Q^2}\right\}
   +\left\{2-(1-\xi)\frac{Q_{\perp}^2}{Q^2}\right\}x\Biggl].
   \label{inx}
\end{align}
\end{widetext}
This result corresponds with the Eq.(8) in the reference \cite{rojas}.
Following \cite{rojas} we get for arbitrary gauge fixing parameter the Eq.~(\ref{allgauge}),
\begin{widetext}
 \begin{align}\label{allgauge}
&\mathcal{M}(y)=m_{0}+\frac{e^2}{4}\int_{0}^{\frac{\Lambda^2}{4\gamma}}\frac{dz}{(2\pi)^2}\mathcal{M}[z]\int_{0}^{1}dxx^{\lambda}
\bigg\{ \notag\\
&\times\bigg[\left\{2(1+x)-(1-\xi)\left(x+(1-x)^2 z-y(1-x)^2+\frac{1}{2}yz(1-x)^3 \right)\right\}\notag\\
&\times e^{y(1-x)}\Gamma[0,y(1-x)]
+(1-\xi)(1-x)\left(\frac{z}{2y}-1+\frac{1}{2}z(1-x)\right)
\bigg]\theta\left(y-z\right)+\notag\\
&\bigg[\left\{2(1+x)-(1-\xi)\left(x+\frac{1}{2}z^2(1-x)^3\right)\right\}\notag\\
&\times e^{z(1-x)}\Gamma[0,z(1-x)]
+(1-\xi)(1-x)\left(-\frac{1}{2}+\frac{1}{2}z(1-x)\right)
\bigg]\theta\left(z-y\right)\bigg\},
\end{align}
\end{widetext}
where $z=Q^2$, $y=p^2/4\gamma$ and
$\lambda=z+\frac{\mathcal{M}^2[\qv]-i\epsilon}{2eB}-1$ and  $\Gamma(x,y)$ is the incomplete gamma function. 
The set of approximations   leading to Eq.~(\ref{allgauge}) are the same that in~\cite{rojas} which  make no reference to the strength of the magnetic field, therefore this equation is valid for arbitrary magnetic field intensities.
 If we take, $\xi=1$, in Eq.~(\ref{allgauge}) it reproduces  Eq. (11) in~\cite{rojas} with $m_0=0$.
For strong magnetic fields it will be useful to compare the solutions to this equation with the solutions tho the corresponding  equation for the lowest Landau level (LLL)
\begin{align}\label{lll}
\mathcal{M}(p_\parallel)=\frac{e^2}{2(2\pi)^2}\Biggl\{\int_{0}^{p^2} dq_{\parallel}^2\frac{\mathcal{M} _A(q_\parallel)}{q_\parallel^2+\mathcal{M} ^2(q_\parallel)}e^{\frac{p^2_{\parallel}}{4\gamma}}\Gamma[0,\frac{p^2_{\parallel}}{4\gamma}]\notag\\
+\int_{p^2}^{\infty} dq_{\parallel}^2\frac{\mathcal{M} (q_\parallel)}{q_\parallel^2+\mathcal{M} ^2(q_\parallel)}e^{\frac{q^2_{\parallel}}{4\gamma}}\Gamma[0,\frac{q^2_{\parallel}}{4\gamma}]\Biggr\}
\end{align}
to calculate this expression we have used the same assumptions of the kernel softness respect to the momentum variables that in Eq.~(\ref{allgauge}).
We can find numerical solutions to the integral equation  Eq.~(\ref{allgauge}) for several magnetic field intensities making an appropriate choice of the cutoff in every range.
For weak magnetic fields,  $eB<m_0^2$,  we take as cutoff  the electron mass given that it  corresponds to the highest momentum  scale and make $\xi=0$. For weak magnetic fields the dimensional reduction is missing and thus it is important to take into account  the  ultraviolet divergences  since we are in a  similar regime  to the vacuum. This fact makes the numerical work harder and conceptually some care should be taken.
There,  contrary  to the  strong magnetic field case,  the 
difference between the mass function and the bare electron mass $m_0$ is meaningless because it does not tend to zero when the magnetic field is turn off, leading to  to a fictitious dynamical mass. We can make an rough estimate of this contribution by consider the QED one loop contribution to the self-energy in the perturbative case. In the Landau gauge we get
\begin{equation}
\delta m=m_{\text{phys}}-m_{0}\sim \frac{3\alpha m_0 }{4\pi}
\ln\frac{\Lambda^2+m_0^2}{m_0^2}\lvert_{\Lambda=m_0}\sim10^{-3}m_0,
\end{equation}
which is  at least two magnitude orders higher than the magnetic field contribution.
 To remove this additive constant is sufficient   consider only differences between the  mass function.
\begin{figure}[htb]
\includegraphics[bb=0  0  15cm 6cm]{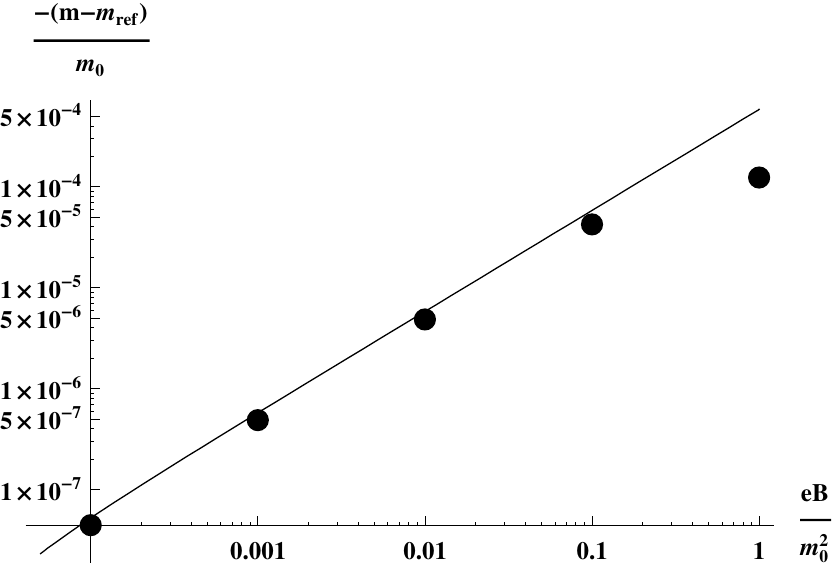}
\caption{Dynamical electron mass as a function of the magnetic field in units of $m_0^2$. The continuous line  
corresponds to the linear contribution Eq.~(\ref{debil}).  Dots  correspond to the numerical solution including ALL Eq.~(\ref{allgauge}). Here $m_{\text{ref}}=m(eB=10^{-5}m_0^2)\sim 2.4\times10^{-3}m_0$}
\label{grafica3}
\end{figure}
Numerical results  are shown in the fig.~\ref{grafica3}. It is noticeable that the relation, Eq.~(\ref{debil}), remains as  the leading term  even for  field intensities close to $B_c =m_0^2/e$. This is an  important result because, despite that the old formula Eq.~(\ref{debil}) is very well established, to the best of our knowledge no validation of this result including higher Landau levels is  known. Besides  fig.~\ref{grafica3} shows that the 
assumption $eB<< m^2$ is no too  restrictive.
It is very well known that  the anomalous magnetic moment of  the electron $\Delta\mu=\frac{\alpha e_0\hbar}{4\pi m_0c}$ can be obtained from the real part of the mass operator which depends of the magnetic field ~\cite{Schwinger,sokolov,raymond,baier,gepraegs,elmfors,Ferrer4}. In our approach we can read it from  fig.~\ref{grafica3} by use the formula  $\Delta\mu=\lvert \delta m(B)/B \rvert$, where  $\delta m(B)$ is the magnetic correction to the electron mass.  We can see in  fig.~\ref{moment} that the anomalous magnetic moment is independent with the magnetic field up to magnetic fields intensities near  $0.1 B_c$ in agreement with previous calculations which only include  lowest Landau levels~\cite{gepraegs}. For magnetic fields near  $B_c$ the anomalous magnetic field is highly suppressed  and the same concept of  magnetic moment in this range should be revisited. In the fig.~\ref{moment}
there is an  small gap between the analytical value and our numerical solution, this should only be a numerical error.
\begin{figure}
\includegraphics[bb=0  0  15cm 5cm]{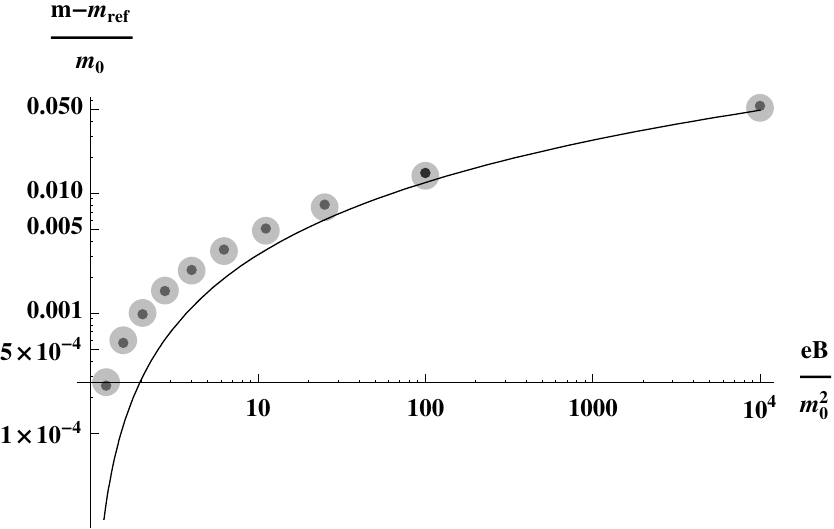}
\caption{Dynamical electron mass as a function of the magnetic field in units of $m_0^2$. The continuous line  
corresponds to the analytical solution of Gusynin and Smilga. Gray and  Black dots  correspond to the numerical solution including ALL and the LLL respectively.  Here $m_{\text{ref}}=m(B=m_0^2/e)$}
\label{grafica1}
\end{figure}
For the strong regime we take $\Lambda=\sqrt{eB}$ and calculate the solutions to Eq.~(\ref{allgauge}) for the interval, $m_{0}^2< eB< 10^4m_{0}^2$, which corresponds to magnetic fields in the phenomenological range . In this regime we make the calculation including higher Landau levels and compare with the LLL solution.  
The solutions for  $eB/m_{0}^2=1$ for ALL and LLL  are $m_{\text{ref}}=2.2\times 10^{-3}m_0$ and $m_{\text{ref}}=1.4\times 10^{-3}m_0$ respectively, shown that high Landau levels represent an important contributions at this field intensities. However if, as in the weak magnetic case, we only consider mass differences  the results are indistinguishable as is shown  in fig.~\ref{grafica1}. This lead us to conclude that for this regime of  coupling, the main effect of higher Landau levels is to include vacuum contributions without  changing significantly the dependence of the mass with the magnetic field.
\begin{figure}[htb]
\includegraphics[bb=0  0  15cm 6cm]{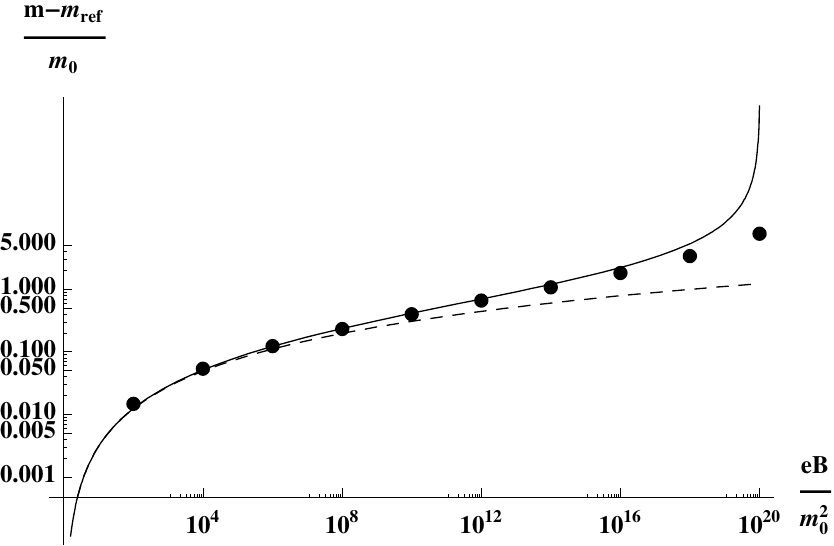}
\caption{Dynamical electron mass as a function of the magnetic field in units of $m_0^2$. The continuous line  
corresponds to the non-perturbative analytical solution of Gusynin and Smilga.  Black dots  correspond to the numerical solution to Eq.~(\ref{lll}) and  the dashed line corresponds to perturbative solution in the Ref.~\cite{jancovici}. Here $m_{\text{ref}}=m(B=m_0^2/e)$}
\label{grafica2}
\end{figure}
For higher magnetic fields we recover the agreement between the lowest Landau level solution and the complete solution. Given  that there are not important contributions by including higher Landau levels for strong magnetic field, we calculate the enhancement of the electron mass and compare the solutions with the previous solutions founded in the literature~\cite{gusyninreal}. From the fig.~\ref{grafica2} we can conclude that the dynamical effects are well described by the perturbative solutions which does not take into account the mass dependence with the momenta. We note the excellent agreement between  the analytical solution in  reference~\cite{gusyninreal} and our numerical solution. This result should not be surprising, since  despite that in reference~\cite{gusyninreal} they begin calculating perturbative corrections, at the end  they get a self-consistent equation for the dynamical mass  and hence  non-perturbative effects are expected in their  approach.
In summary: we have shown that the Eq.~(\ref{allgauge}) can reproduce perturbative as non-perturbative results for different ranges in the magnetic field intensity.  This equation is easily solved with  lower computational power, allowing to carry out   calculations for magnetic fields relevant for the typical astrophysical conditions. 
To avoid misleading conclusions when dealing numerically with higher Landau levels, we have  shown the importance of distinguishing  vacuum contributions from  magnetic field contributions.
Numerically this can be easily implemented by only consider differences between the mass function for  different magnetic field intensities.
Furthermore, in this work we have shown that the magnetic independence of the anomalous magnetic moment is consistent  up to
magnetic field intensities near  $10^{12}$Gauss. At same time, our technique allows to make a precise description of the way in which this assumption is broken in the magnetic field interval $10^{12}-10^{13}$Gauss. 
A natural extension  of this work is to consider  magnetic field intensities in the interval $10^{0}-10^{8}$Gauss. For this an improvement of our numerical techniques is needed. Due to highly precise measurements of the  anomalous magnetic moment non-perturbative effects are  strongly bounded. If we extrapolate  our calculations
to magnetic field intensities of $10^{5}$Gauss, where  some experiments currently
take  place, it is not completely clear that this contributions become less than one part in $10^{12}$ which represents the nowadays precision in the anomalous magnetic moment.  This deserves a future investigation and can be an excellent test for non-perturbative techniques.

The author acknowledge   V. Gusynin, E. Ferrer, V. de la Incera, G. Toledo and W. F. Cuervo for   useful comments  and suggestions, and  A. Ayala, A. Bashir and A. Raya for introducing me to this subject. Financial support for this work has been received of DGEP-UNAM. 
. 

\nocite{clave}
\bibliographystyle{unsrt}
\bibliography{bibliografia}

\end{document}